# Multimode vibrational coupling across the insulator-to-metal transition in 1T-TaS$_2$ in THz cavities

Giacomo Jarc[1,2,3], Shahla Yasmin Mathengattil[1,2], Angela Montanaro[1,2,3], Enrico Maria Rigoni[1,2,3], Simone Dal Zilio[4], and Daniele Fausti[1,2,3,a)].

[1]*Department of Physics, Università degli Studi di Trieste, 34127 Trieste , Italy*
[2]*Elettra Sincrotrone Trieste S.C.p.A., 34127 Basovizza, Trieste, Italy*
[3]*Department of Physics, University of Erlangen-Nürnberg, 91058 Erlangen, Germany*
[4]*CNR-IOM TASC Laboratory, 34139 Trieste, Italy*

a) *Correspondence to: daniele.fausti@fau.de*

**The use of optical cavities on resonance with material excitations allows controlling light-matter interaction in both the regimes of weak and strong coupling. We study here the multimode vibrational coupling of low energy phonons in the charge-density-wave material 1T-TaS$_2$ across its insulator-to-metal phase transition. For this purpose, we embed 1T-TaS$_2$ into THz Fabry-Pérot cryogenic cavities tunable in frequency within the spectral range of the vibrational modes of the insulating phase and track the linear response of the coupled phonons across the insulator-to-metal transition. In the low temperature dielectric state we reveal the signatures of a multimode vibrational strong collective coupling. The observed polariton modes inherit character from all the vibrational resonances as a consequence of the cavity-mediated hybridization. We reveal that the vibrational strong collective coupling is suppressed across the insulator-to-metal transition as a consequence of the phonon-screening induced by the free charges. Our findings emphasize how the response of cavity-coupled vibrations can be modified by the presence of free charges, uncovering a new direction toward the tuning of coherent light-matter interaction in cavity-confined correlated materials.**

## I. INTRODUCTION

Strong and weak coupling regimes between confined light fields and solid-state excitations are attracting increasing attention owing to the potential that such coupling regimes offer to control energies and dissipative properties of materials [1-5]. While in the weak coupling regime, only the dissipative properties of the targeted material are modified [6-8], if the optical confinement is sufficiently strong, i.e. in the strong coupling limit, the light-matter interaction within the cavity overcomes the dissipative processes occurring in the bare systems, and the wave-functions of the material and photon inside the cavity are coherently mixed. This coherent superposition between the material and cavity excitations results in the formation of hybrid light-matter states called polaritons [3, 9, 10], which have different eigen-energies with respect to the uncoupled resonances and carry the spectral weight of all the material's emitters within the cavity mode volume. It has recently been demonstrated that confined electric fields in the strong and weak coupling regimes can modify material functionalities [1, 2], such as the magneto-transport in two-dimensional gases [11], the topological protection of the integer and fractional quantum Hall effect [12, 13], and the ferromagnetic order in an unconventional superconductor [14], or mediate a thermal control of the metal-to-charge density wave insulator transition in 1T-TaS$_2$ [15].

In particular, when photons are hybridized with specific vibrational modes, the physical and optical properties of those vibrations, such Raman scattering dynamics [16], excited state lifetimes [17], emissivity [18], and phonon non-linearities [19], can be altered. It has been demonstrated in the past years that the formation of such vibro-polaritonic states allows for the manipulation of molecular processes, such as chemical reactivity [20-22], molecular and crystal structure [23], and charge transport [24-27].

Here, we study the multimode vibrational coupling of low-energy phonons in the charge-density-wave material 1T-TaS$_2$ and investigate how the change in the charge transport across the 1T-TaS$_2$ insulator-to-metal transition affects the electrodynamics of the cavity-coupled vibrations. For this purpose we embed 1T-TaS$_2$ into THz Fabry-Pérot cavities tunable in frequency at cryogenic temperatures [28] (Fig. 1A) and exploit THz spectroscopy to track the linear response of the coupled-phonons across the insulator-to-metal transition. Coupling between multiple vibrational resonances and a resonant optical cavity have been already demonstrated in molecular frameworks where many sharp vibrational transitions can exist within a narrow spectral window [29-31]. However, no studies of multimode vibrational collective coupling have been reported in condensed-matter complex systems, where phonons can couple to free charges or bound electrons, inducing cooperative phases such as superconductivity, charge-density waves, ferroelectricity or ferromagnetism. Intriguingly, we show here that in the complex material 1T-TaS$_2$, the presence of free charges can effectively modulate the light-phonon collective coupling of the cavity-confined material, demonstrating a previously unexplored path for engineering the light-matter interaction strength [32-36] in a cavity-confined system.

1T-TaS$_2$ is a correlated Transition Metal Dichalcogenide (TMD), extensively studied as one of the first quasi-2D materials in which a charge-density-wave (CDW) order was observed [37]. 1T-TaS$_2$ exhibits various thermodynamical ground states at different temperatures, as well as fertile out-of-equilibrium hidden phases [38-39]. The complex phase diagram of 1T-TaS$_2$, which originates from the intricate competition of Coulomb repulsion, interlayer hopping, lattice strain, and Fermi surface nesting [39-42], exhibits four thermodynamical phases characterized by significant differences in the charge order and mobility [40], and in the commensuration of the CDW.

At very high temperatures (T > 550 K), 1T-TaS$_2$ presents the features of a simple metal and no charge-density-wave is present. As temperature is decreased below 550 K, an incommensurate charge-density-wave (IC-CDW) appears, transitioning to a nearly-commensurate CDW (NC-CDW) at 350 K. The NC-CDW phase is characterized by commensurately ordered hexagonal-shaped polarons separated by conductive domain boundaries [39, 43]. Within the commensurate domain-like regions, the lattice distortion creates "star-of-David" polaron clusters [44-46], consisting of a single electron localized on a central Ta atom and surrounded by 12 Ta atoms displaced toward the central one (top inset of Fig. 1B).

Below 180 K, a transition to an insulating phase occurs. Below 180 K, the metallic domain boundaries, related to the discommensuration network of the CDW regions, disappear and the system becomes a fully commensurate charge-density-wave (C-CDW) insulator [41]. The illustrations of the melting of the discommensurate CDW regions, associated with the metal-to-insulator charge ordering transition studied in this work, are presented in the insets of Fig. 1B.

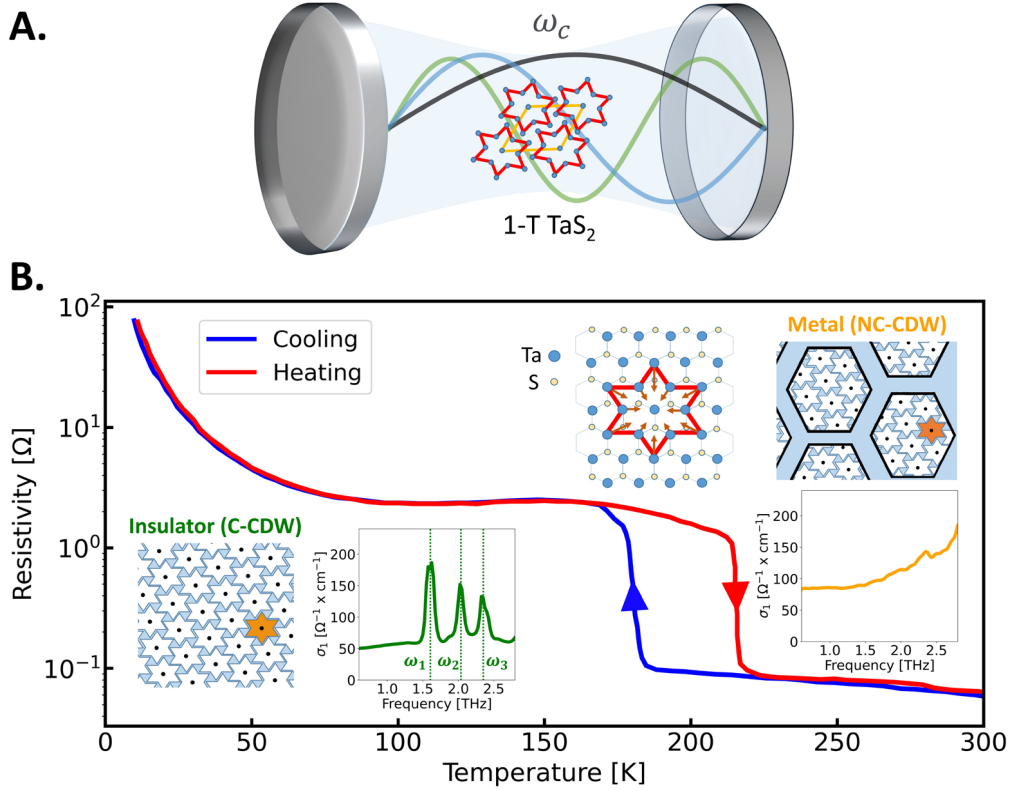

**Figure 1**: **Experimental configuration. A.** Sketch of 1T-TaS$_2$ embedded in the center of a cryogenic optical cavity with fundamental mode frequency $\omega_c$ tunable in the THz range. **B.** Temperature dependence of the in-plane resistivity measured upon cooling and heating (adapted from Ref. [41]), marking the metal-to-insulator transition and its hysteresis. In the insets, the THz optical conductivity $\sigma_1$ measured in the insulating (green) and metallic (orange) states is shown, together with the sketch of the in-plane lattice modulations characteristic of the insulating C-CDW phase and of the metallic NC-CDW phase.

Importantly, while in the nearly-commensurate CDW phase the metallic response is determined by the charged domain boundaries, below the critical temperature (180 K), in the commensurate CDW state, the discommensuration network melts, giving rise to a frozen charge-density-wave with an insulating response. The main panel of Fig. 1B shows the temperature-dependent resistivity (adapted from Ref. [41]) of a bulk 1T-TaS$_2$ crystal measured across the NC-C metal-to-insulator transition studied in this work. The resistivity shows that the metal-to-insulator transition between the nearly-commensurate CDW and the commensurate CDW is a first order transition characterized by a hysteresis. The phase transition occurs indeed at ~ 180 K (215 K) by cooling (heating) the sample.

The change in the charge transport properties of 1T-TaS$_2$ across the metal-to-insulator transition is mapped in the THz linear response. The optical features associated with the charge-ordering transition, as shown by the THz optical conductivity $\sigma_1(\omega)$ measured in the metallic and dielectric phases (insets of Fig. 1B), are as follows:

i. The increase in the low frequency optical conductivity $\sigma_1(\omega)$ (~ 0.3 THz < $\omega$ < 1.4 THz) above the critical temperature, consistent with a transition to a metallic state. The Drude-like linear response of the free carriers is indeed enhanced in the metallic NC-CDW phase, giving rise to an increased absorption within the quasi-static spectral region [47, 48].

ii.  The emergence below the critical temperature of IR-active phonons at $\omega_1 = 1.58$ THz, $\omega_2 = 2.02$ THz, and $\omega_3 = 2.35$ THz, which are allowed by the symmetry of the fully commensurate CDW. As shown in the $\sigma_1(\omega)$ plots presented in the insets of Fig. 1B, the C-CDW vibrations are screened by the free carriers [47] and therefore not visible in the metallic phase.

## II.  EXPERIMENTAL METHODS

To study the vibrational coupling of the C-CDW vibrations across the insulator-to-metal transition, we embed an ~ 10 μm thick 1T-TaS$_2$ sample within a cryogenic Fabry-Pérot cavity tunable in frequency in the THz range. Further details can be found in [28] and in the Supplementary Material. The cryogenic THz cavity is composed of two cryogenic-cooled piezo-controlled semi-reflecting mirrors between which the sample is inserted. Cavity mirrors are mounted on copper holders, and they are cryo-cooled by means of copper braids directly connected to the cold finger of the cryostat. In order to ensure the tuning of the mirror distance at cryogenic temperatures, the piezo-actuators are thermally decoupled from the mirror supports by means of a PEEK (Poly-Ether Ether Ketone) disk and three ceramic cylinders. The cavity chamber is mounted on a flow cryostat with a temperature feedback circuit enabling temperature scans at a fixed cavity frequency. Temperature is measured on the sample's holder and on the mirror supports for reference. The procedure employed to correct the thermal drifts of the cavity resonance and hence ensure a temperature scan at a fixed cavity length is presented in the Supplementary Material.

Cavity semi-reflecting mirrors are fabricated by evaporating a thin bilayer of titanium-gold (2-15 nm) on a 2 mm thick crystalline quartz substrate. We measure the transmission amplitude of a single cavity mirror to be ~ 15 % in the THz spectral range of the experiment with no apparent spectral features. For the employed thickness of the gold layer (15 nm), the Fabry-Pérot cavity has a quality factor Q ~ 6, which does not significantly depend on the temperature of the cavity (see the Supplementary Material).

The ~ 10 μm-thick single crystal 1T-TaS$_2$ sample is mounted within the two mirrors in a copper sample holder directly connected to the cold finger of the cryostat and sealed between two 2 μm-thick silicon nitride (Si$_3$N$_4$) membranes. Importantly, the membranes do not show any spectral dependence in the THz spectral range employed in the experiment.

We employ broadband THz spectroscopy to characterize the vibrational coupling of the C-CDW modes with the cavity field across the insulator-to-metal phase transition. Nearly single-cycle THz pulses are generated via the acceleration of the photoinduced carriers in a large-area GaAs-based photoconductive antenna (PCA) [49]. The photoexcitation is achieved by pumping the PCA with an ultrashort laser pulse (50 fs pulse duration, 745 nm central wavelength). The emitted collimated THz beam is then focused on the sample mounted inside the cavity. The field transmitted through the sample-cavity assembly is probed by standard Electro-Optic-Sampling (EOS) in a 0.5 mm ZnTe crystal. The employed THz fields have a broad spectral content, reaching frequencies up to ~ 6 THz. (see [28] and the Supplementary Material).

## III. RESULTS AND DISCUSSION

### A. MULTIMODE VIBRATIONAL STRONG COLLECTIVE COUPLING OF THE C-CDW PHONONS

#### 1. THz spectroscopy of the strongly-coupled phonons at 80 K

We study here the strong collective coupling of the multimode phonon structure of the dielectric C-CDW phase with the single-mode cavity field. We expect the strong coupling between the fundamental cavity mode and the three IR-active vibrations of the CDW to generate 3 + 1 = 4 non-degenerate polariton modes: an upper (UP) and a lower (LP) polariton, together with two middle polariton states (P1, and P2), resulting from the photon-mediated hybridization of the C-CDW phonons. Note that, as in the single phonon configuration, the cavity-mediated hybridization is associated with the formation of dark states (DS) at the frequency of each bare vibrational resonance. These states are dipole-not-allowed transitions and hence invisible to linear spectroscopy. The scheme of the energy levels resulting from the hybridization of the multiple C-CDW vibrations with the single cavity resonance is sketched in Fig. 2A.

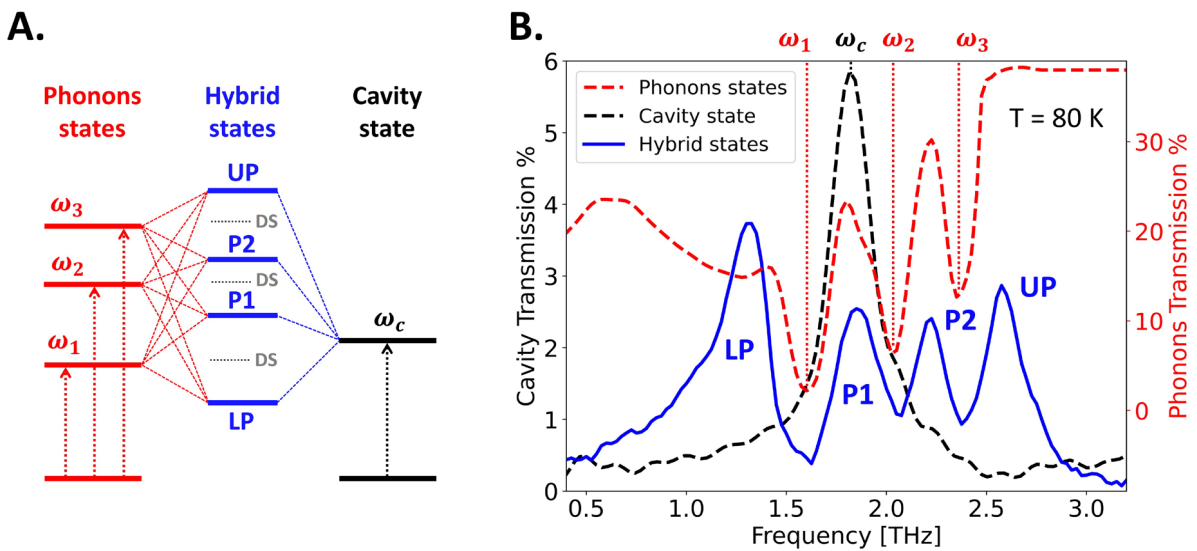

**Figure 2**: **Multi-polariton hybridization of the C-CDW phonons of bulk 1T-TaS$_2$. A.** Energy level scheme of the multi-polariton hybridization of the C-CDW phonons. Strong collective coupling between the three C-CDW phonons at frequencies $\omega_1$, $\omega_2$, and $\omega_3$ and the single cavity mode at frequency $\omega_c$ results in the formation of four non-degenerate hybrid states: an upper (UP) and a lower (LP) polariton, and two middle polariton resonances (P1, and P2). Dark states resulting from the strong coupling are denoted with DS. **B.** Transmission spectrum (blue, left axis) of bulk 1T-TaS$_2$ at 80 K within a THz cavity such that the frequency $\omega_c$ of the optical mode (black dashed spectrum, left axis) is at the midpoint of two C-CDW phonons. The upper (UP) and lower (LP) polariton states, together with the two middle polariton modes (P1, and P2) are visible in the THz transmission spectrum. The spectrum of the uncoupled C-CDW modes at 80 K is shown in red on the right axis for reference.

Fig. 2B presents a representative THz transmission spectrum of the polariton modes measured on the bulk 1T-TaS$_2$ crystal. The spectrum has been obtained at 80 K in a representative cavity configuration in which the cavity fundamental mode (black dashed line) resonates at the midpoint of two adjacent CDW phonons (red dashed line). We note that bright peaks associated with the lower (LP), upper (UP), and middle (P1, and P2) polariton states are visible in the THz linear transmission. Importantly, the Rabi splittings between adjacent polaritons estimated from the measured transmission spectrum ($\Omega_{ij}$ = 0.496, 0.301, 0.310 THz) exceed both the bare phonon linewidths ($\gamma^{phon}_{1,2,3}$ = 0.172, 0.154, 0.148

THz) and the bare cavity linewidth ($\gamma_{cav}$ = 0.290 THz). This evidence implies that the rate of coherent energy exchange between the confined electromagnetic field and the C-CDW phonons, which is proportional to $1/\Omega_{ij}$, dominates over the bare dissipative dynamics, allowing Rabi oscillations to occur in the studied system.

In order to shed light on the dispersive properties of the measured multimode vibro-polaritons, we tuned the cavity frequency at 80 K and measured, for each cavity fundamental mode, the THz transmission of the strongly-coupled system. Fig. 3A presents the THz transmission for each cavity frequency across the phonon resonances and the obtained dispersion of the multimode polaritons. We highlight two different behaviors as a function of the cavity frequency:

i. The lower and upper polaritons (UP, and LP) strongly disperse upon tuning the cavity fundamental mode. In particular, their energies approach the energy of the bare modes (uncoupled cavity and C-CDW phonons) under off-resonance conditions.

iii. The frequencies of the two middle polaritons (P1, and P2) do not exhibit a significant dependence on the cavity frequency, and they lay in a dark region of the vibrational spectrum, where the C-CDW vibrations are not allowed for the sample in free space (Fig. 2B).

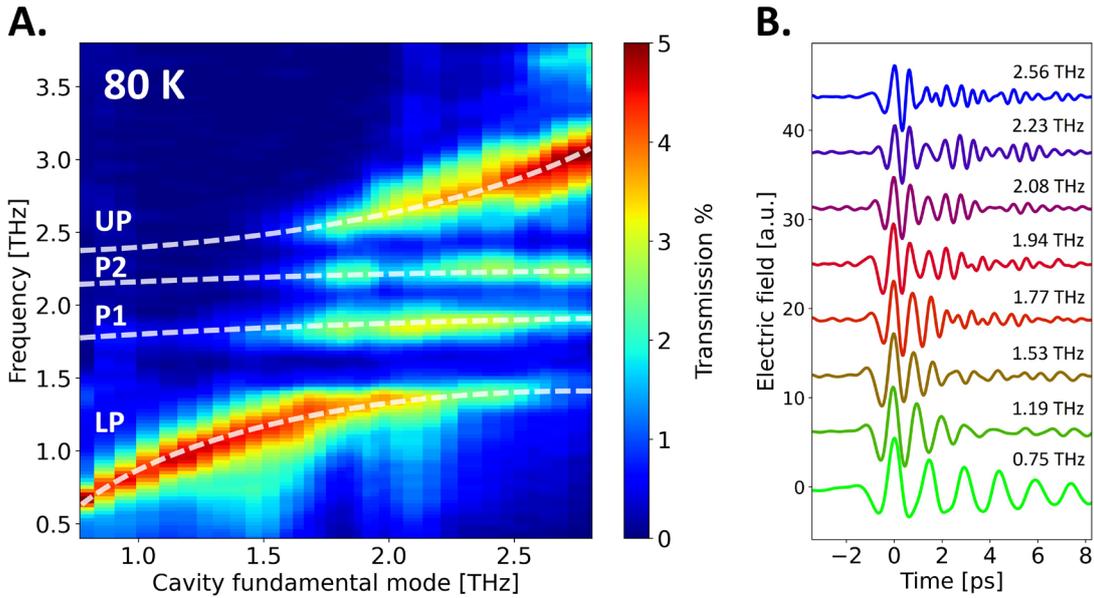

**Figure 3**: **Multimode polariton dispersion in the C-CDW phase of bulk 1T-TaS$_2$**. **A.** Dispersion of the four polariton branches measured in the C-CDW phase at 80 K. The lower (LP) and upper (UP) polariton energies strongly disperse as a function of the cavity frequency. On the contrary, no significant dispersion is measured for the two middle polariton states (P1, and P2). **B.** THz time-domain fields exiting the sample-cavity assembly at 80 K at different cavity frequencies (indicated in the legend) across the phonon energies. The measured THz electric fields have been spectrally filtered in the range 0.35 – 3.8 THz. The THz time-domain traces have been vertically shifted for clarity.

The multimode polariton structure can be also visualized by looking at the THz fields measured at the output of the sample-cavity assembly (Fig. 3B). At low cavity frequencies, i.e., off-resonance with the C-CDW modes, the THz fields present indeed a single-decaying oscillatory dynamics characteristic of a single cavity mode. Conversely, when the cavity fundamental mode approaches the spectral region of the C-CDW vibrations, multiple Rabi oscillations are observed. We highlight that, due to the multimode nature of the light-phonon coupling, we do not observe a single beating modulation of the decaying electric field, as in the single mode case [28, 50, 51]. The multiple modulation structure originates from the fact that, in the studied system, the cavity photons are at the same time coherently exchanging energy with all the three non-degenerate vibrational modes.

## 2. Multimode Dicke model

In order to relate the different dispersive properties of the polaritons (Fig. 3A) to quantum superposition effects between the bare C-CDW phonons and the cavity field, we resort to a coupled-oscillator Dicke model in the single photon case [3, 52]. The full Hamiltonian $\widehat{H}$ describing the strong collective coupling between the three C-CDW excitations and the single cavity mode can be written as:

$$\widehat{H} = \widehat{H}_{cav} + \widehat{H}_{phon} + \widehat{H}_{int} = \begin{pmatrix} \omega_c & \frac{V_1}{2} & \frac{V_2}{2} & \frac{V_3}{2} \\ \frac{V_1}{2} & \omega_1 & 0 & 0 \\ \frac{V_2}{2} & 0 & \omega_2 & 0 \\ \frac{V_3}{2} & 0 & 0 & \omega_3 \end{pmatrix}. \qquad (1)$$

The first term $\widehat{H}_{cav}$ describes the uncoupled cavity oscillator, having tunable frequency $\omega_c$. The second term $\widehat{H}_{phon}$ models the three uncoupled C-CDW phonons, having bare frequencies $\omega_{1,2,3}$. $\widehat{H}_{int}$ describes the cavity-phonon interaction in the rotating wave approximation. The collective coupling strengths of the C-CDW modes with the cavity field are quantified by the light-phonon interaction energies $V_{i(i=1,2,3)}$. The phonon collective-couplings with the cavity field inside the Hamiltonian ($V_i$) can be experimentally estimated by the Rabi splittings $\Omega_{ij}$ [3], which correspond to the minimum energy separation between adjacent polaritons.

Importantly, the coupled Hamiltonian in Eq. 1 allows to quantify the composition of a given polariton wave-function in terms of a linear combination of the original, uncoupled resonances. Supposing to have within the cavity mode volume $N$ excited phonons for each of the three measured C-CDW modes, each polariton wave-function $|\psi_{PL}\rangle$ can be expressed on the basis of the uncoupled cavity/phonons modes as [53]:

$$|\psi_{PL}\rangle = X_{cav}(\omega_c)|0,0,0;1\rangle + $$
$$+ X_1(\omega_c) \sum_{i=1}^{N} |e_i,0,0;0\rangle + X_2(\omega_c) \sum_{i=1}^{N} |0,e_i,0;0\rangle + X_3(\omega_c) \sum_{i=1}^{N} |0,0,e_i;0\rangle. \qquad (2)$$

Here, we have indicated with $|0,0,0;1\rangle$ the purely cavity state, and with $\sum_{i=1}^{N} |e_i,0,0;0\rangle$, $\sum_{i=1}^{N} |0,e_i,0;0\rangle$, and $\sum_{i=1}^{N} |0,0,e_i;0\rangle$ the purely vibrational states in which respectively the first, second, and third CDW phonons are in their excited states $|e_i\rangle$. $X_{cav}(\omega_c)$ denotes the cavity fraction of the hybrid polariton wave-function $|\psi_{PL}\rangle$, while $X_{1,2,3}(\omega_c)$ denote the phononic contribution given, respectively, by the first, second, and third C-CDW mode to $|\psi_{PL}\rangle$.

We note that by experimentally tuning the cavity length, the dispersive term $\omega_c$ of the coupled Hamiltonian $\widehat{H}$ can be tuned through the non-dispersive C-CDW modes to control the eigen-energies of the phonon-polaritons. This will result in the anti-crossing multi-polariton branches experimentally observed and reported in Fig. 3. In order to show that the model is suitable for reproducing the energy splittings among the cavity-coupled C-CDW hybrid modes, we calculate the dispersion of the polariton frequencies as a function of $\omega_c$ by diagonalizing $\hat{H}$ (Eq. 1) and use them to fit the polariton frequencies measured experimentally (Fig. 3A). We use as fit parameters the three interaction energies $V_i$ of the three non-degenerate phonon modes. The result of the fitting procedure is presented in Fig. 4A. The fitted cavity-phonon collective couplings ($V_{1,2,3}$ = 0.457, 0.287, 0.291 THz) are in agreement with the Rabi splittings ($\Omega_{ij}$ = 0.496, 0.301, 0.310 THz) estimated from the measured polariton dispersion (Fig. 3A). This shows the applicability of the coupled Hamiltonian (Eq. 1) to the experimental setting.

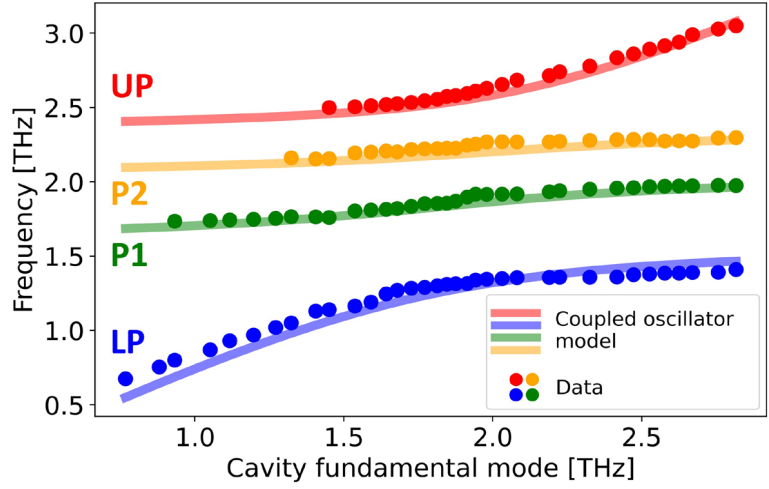

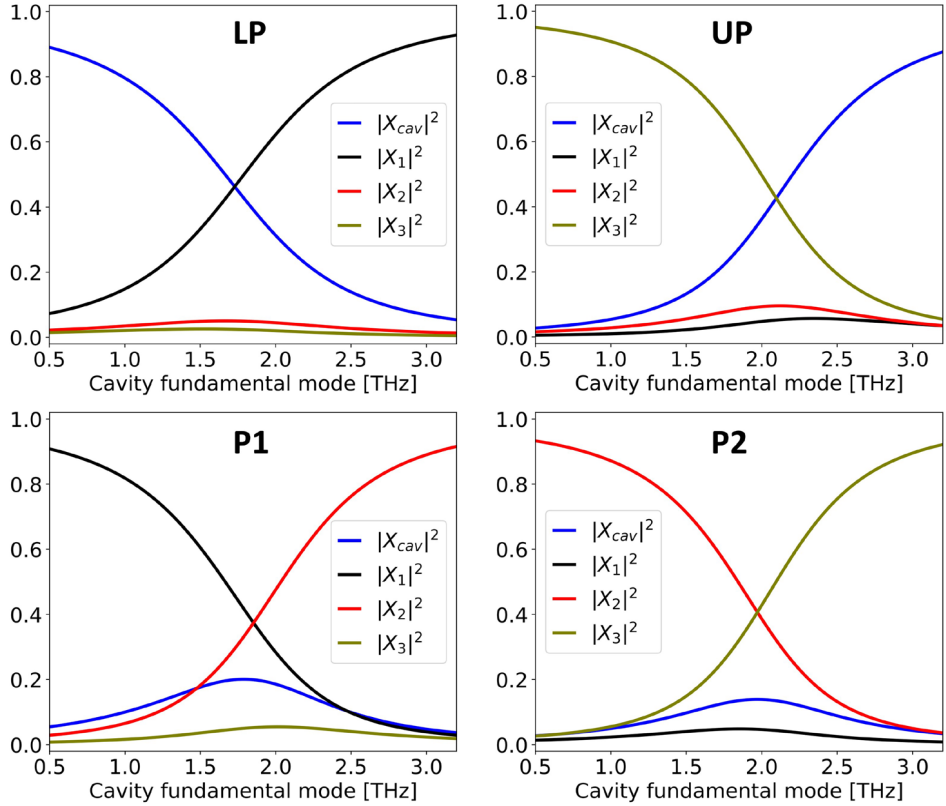

**Figure 4**: **Estimated wave-function components of the polariton states of the C-CDW phase**. **A.** THz vibro-polariton branches for the four light-matter hybrid states (LP, P1, P2, and UP). The filled circles correspond to the measured polariton peaks at 80 K. The curves represent the eigen-energies obtained from the coupled oscillator model fitted to the measured polariton peaks. Note that only the polariton peaks above the noise level have been fitted. **B.** Cavity ($|X_{cav}|^2$) and phonon ($|X_{1,2,3}|^2$) fractions of the polariton wave-functions as a function of the cavity fundamental mode $\omega_c$. The estimations of the wave-function fractions are based on the coupled oscillator model (Eq. 1).

The evolution of the cavity/phonon fractions ($X_{cav}/X_{1,2,3}$) of the four polaritonic wave-functions upon tuning the cavity frequency $\omega_c$ is presented in Fig. 4B. The wave-function fraction for each polariton resonance has been estimated by diagonalizing $\widehat{H}$ (Eq. 1) on the multimode cavity-phonon basis (Eq. 2) with in input the bare phonon frequencies ($\omega_{1,2,3}$ = 1.58, 2.02, 2.35 THz) and the light-phonon interaction energies fitted from the experimental dispersion ($V_{1,2,3}$ = 0.457, 0.287, 0.291 THz). The extracted coefficients of the wave-functions show that the four polaritons have character from all the three C-CDW modes as a result of the cavity-mediated hybridization. Crucially, we highlight distinct behaviors between the upper (lower) polariton states and the middle polariton resonances.

i. The lower and upper polaritons (UP, and LP) exhibit a strong cavity component $|X_{cav}|^2$, which justifies their strongly dispersive properties reported in the transmission spectra of Fig. 3A. In 'particular, the anti-crossing behaviour of the lower polariton branch originates mainly from the hybridization of the C-CDW phonon at frequency $\omega_1$ with the cavity field (Fig. 4B). Conversely, the upper polariton wave-function is dominated by the coherent mixing of the C-CDW mode at frequency $\omega_3$ and the single cavity mode (Fig. 4B). Note that, since the Rabi splittings $\Omega_{ij}$ are smaller with respect to the frequency separation of the uncoupled phonons, a uniform mixing among all the three C-CDW vibrations is inhibited.

ii. The middle polaritons (P1, and P2) wave-functions display a substantially lower cavity fraction $|X_{cav}|^2$. This evidence justifies their non-dispersive behaviour upon tuning the cavity fundamental frequency. The coupled oscillator model (Eq. 1) indicates that the middle polaritons have a strong phononic component, which results from the cavity-mediated hybridization of different non-degenerate C-CDW excitations. In particular, as shown in Fig. 4B, the P1 polariton mainly results from the coherent superposition of the $\omega_1$ and the $\omega_2$ phonons, while the P2 polariton results from the hybridization between the $\omega_2$ and the $\omega_3$ vibrational modes.

### B. TEMPERATURE DEPENDENCE OF THE VIBRO-POLARITON MODES

Proven that a multimode vibrational strong collective coupling can be established at low temperatures, we study here how the change in the conductive properties of 1T-TaS$_2$ across the insulator-to-metal transition affects the response of the strongly-coupled C-CDW phonons.

Figure 5A presents the dispersion of the coupled cavity measured at different temperatures across the insulator-to-metal transition. Dispersions have been measured by heating the sample from the insulating C-CDW state, where the CDW phonons are strongly coupled to the single cavity mode (Fig. 3). The temperature dependence of the cavity dispersion presented in Fig. 5A shows that, upon increasing the temperature, the Rabi splittings between the multimode polaritons, which are observed in the dielectric phase of 1T-TaS$_2$, collapse across the insulator-to-metal transition. This effect is related to the screening of the C-CDW phonons induced in the metallic phase by the free charges. The presence of free carriers creates indeed an additional scattering channel for the bound C-CDW excitations, increasing their dissipative rates and effectively screening their linear response to the THz electromagnetic field [47, 48] (see the optical conductivities $\sigma_1$ in Fig. 1B).

To more clearly highlight the suppression of the polariton splittings across the insulator-to-metal transition, we present in Fig. 5B the temperature evolution of the THz transmission at a fixed cavity frequency ($\omega_c \sim 2$ THz). The cavity frequency selected for this comparison ($\omega_c \sim 2$ THz) lies within the spectral region of the C-CDW vibrations.

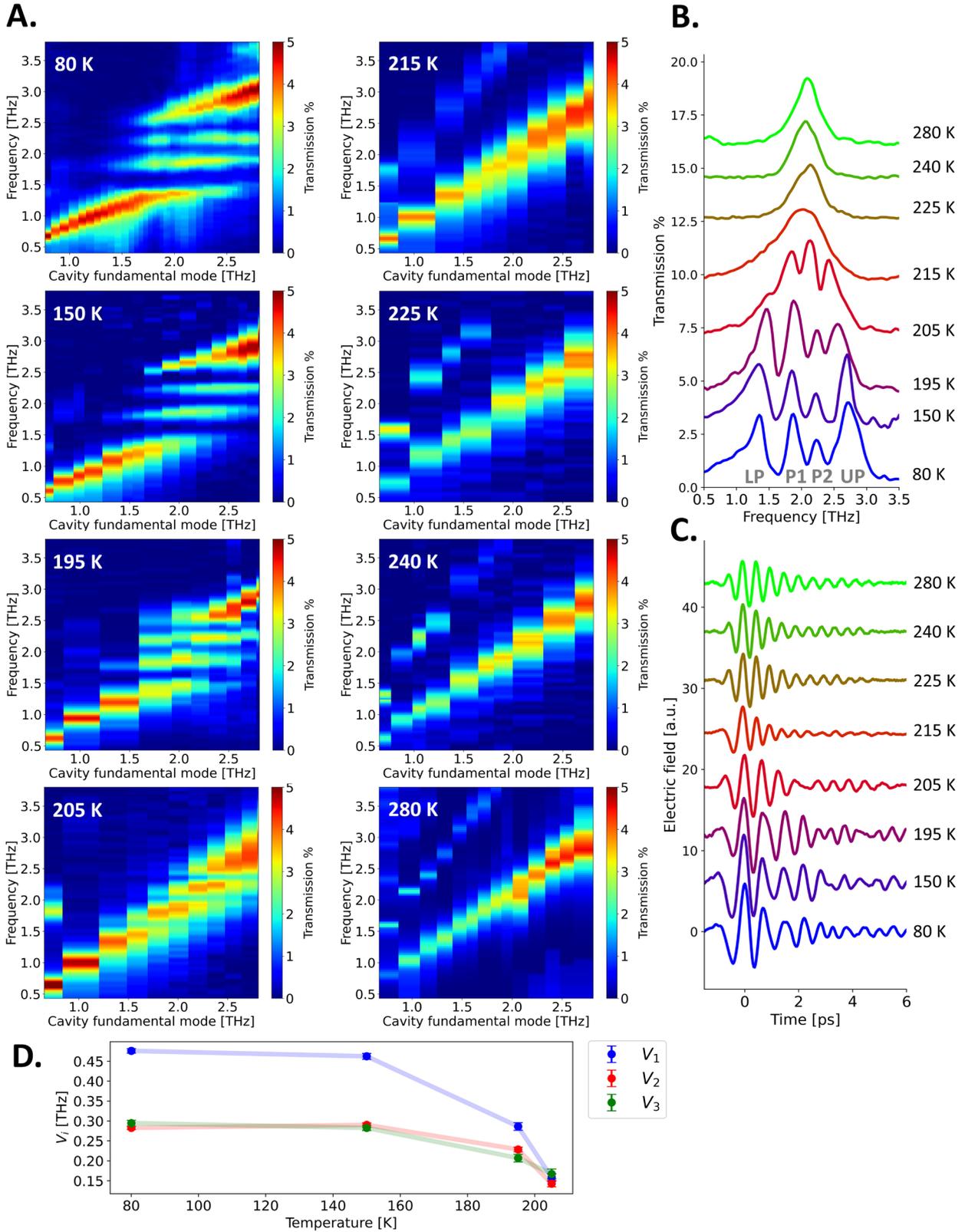

**Figure 5: Temperature dependence of the cavity dispersion measured across the insulator-to-metal transition in 1T-TaS$_2$. A.** Cavity dispersion measured at different temperatures (indicated inside the plots) across the insulator-to-metal phase transition. **B.** Evolution of the THz transmission across the insulator-to-metal transition for a fixed cavity frequency ($\omega_c \sim 2$ THz) within the spectral region of the C-CDW phonons. The spectra have been vertically shifted for clarity.
**C.** THz fields corresponding to the spectra in B. exiting the ~ 2.0 THz cavity at different temperatures. The THz traces have been vertically shifted for clarity. All the temperature-dependent measurements presented in the figure have been measured upon heating the sample within the cavity from the insulating C-CDW phase. **D.** Dependence of the light-phonon interaction energies $V_{i(i=1,2,3)}$ across the insulator-to-metal transition. The interaction energies have been estimated by fitting the polariton dispersions below the nominal Tc (spectra of panel A) with the multimode cavity-phonon Dicke Hamiltonian (Eq. 1).

Fig. 5B highlights that in the metallic phase, where the C-CDW phonons are fully screened by the Drude-like response of free charges, the multimode polariton resonances characteristic of the dielectric state evolve in a single resonance having the frequency of the bare cavity.

The collapse of the polariton resonances in proximity of the insulator-to-metal transition can be also visualized in the time-domain THz fields measured at the output of the coupled cavity. Fig. 5C presents the THz fields exiting the sample-cavity assembly across the phase transition. The THz time-domain fields of Fig. 5C correspond to the THz spectra shown in Fig. 5B. We note that, by increasing the temperature, the free-charge screening induces a suppression of the multiple Rabi oscillations near the critical temperature (Tc ~ 215 K). Above Tc, i.e. in the metallic state, the multiple Rabi oscillations evolve in a single decaying oscillatory response, as expected from the full screening of the C-CDW modes [47].

In order to insigth into how the presence of free charges affects the light-phonon interaction strength inside the cavity, we leverage on the multimode Dicke Hamiltonian (Eq. 1) to track the light-phonon collective couplings $V_i$ across the insulator-to-metal transition. The estimations of $V_i$ have been made approaching Tc from the low-temperature insulating phase, i.e. where the multimode vibro-polaritons are visible (Fig. 5A). The temperature dependence of the coupling energies between the single cavity mode and the three CDW phonons is shown in Fig. 5D. For each temperature, the phonon-cavity collective couplings $V_i$ have been estimated by fitting the maximum of the vibro-polariton branches with the eigen-energies extracted from the multimode Dicke model (Eq. 1). The extended fits for each temperature are presented in the Supplementary Material, Fig. S3. The results of the fits (Fig. 5D) show that the collective coupling of each CDW phonon with the cavity electric field is suppressed in proximity of the transition to the metallic state. This evidence confirms that the onset of metallicity in proximity of the charge ordering transition effectively modulates the light-phonon couplings within the cavity, inducing a suppression of the strong collective light-phonon hybridization observed in the low-temperature dielectric state. We additionally observe that, although the vibro-polariton resonances collapse as the system approaches the metallic state, the individual polaritons (LP, P1, P2, and UP) continue to disperse with the cavity frequency with a trend similar to that observed at 80 K. This suggests that, although the free-charge screening suppresses the cavity-phonon collective couplings, the dependence of the phonon-polariton wave-functions on the cavity detuning, which rule the dispersion of the polaritonic modes, undergoes no substantial modifications in proximity of the insulator-to-metal transition (see Fig. S4 of the Supplementary Material).

Finally, we note that no relevant changes in the effective critical temperature are observed under the collective strong coupling of the C-CDW phonons, conversely to the effects observed with lower frequency cavities, i.e. in the sub-THz region [15]. The effective modification of the phase transition temperature observed in Ref. [15] was rationalized as it is mainly due to a thermal Purcell-like mechanism in which the spectral profile of the cavity modifies the energy exchange between the excitations in the material and the external electromagnetic field, thus modulating the radiative heat load on the sample. It should be stressed that, as discussed in Ref. [54], the set-up employed in Ref. [15] and here is particularly sensitive to the thermal Purcell effect of low frequency cavities (< ~ 2 THz), while it is not sensitive to higher frequencies being the external radiation above ~ 2 THz screened by the "cold filtering" effect from the mirror quartz substrates. For this reason, we cannot exclude the presence of a thermal Purcell-like scenario for a cavity coupled to the C-CDW phonons as the present set-up would not be particularly sensitive to a thermal Purcell-effect in the frequency region of the C-CDW vibrations. Further dedicated studies are necessary to address thermal Purcell effects for cavities coupled to phonons in the mid-IR range.

## IV. CONCLUSIONS

In conclusion, we have studied how the change in the charge-transport properties across the insulator-to-metal transition in 1T-TaS$_2$ affects the light-phonon coupling in the cavity-confined material. We showed that by tuning the cavity mode resonantly to the IR-active phonons of the dielectric C-CDW phase, the THz signatures of a multimode polariton mixing are observed. The estimated components of the polaritonic wave-functions show that the measured polaritons of the C-CDW phase inherit character from all the vibrational resonances within the employed spectral range as a consequence of the cavity-mediated hybridization. In particular, together with an upper and a lower polariton resonance, we observed two weakly-dispersive middle polariton states resulting from the cavity-mediated mixing of non-degenerate phonons of the C-CDW insulating phase.

The Rabi splittings between the vibro-polaritons are suppressed across the insulator-to-metal transition, as a consequence of the screening of the free charges. The free-charge screening induces indeed a modification of the coherent collective coupling between the CDW vibrations and the cavity field in proximity of the insulator-to-metal transition. The evidence that the free-charge screening of bound vibrational excitations can modulate the light-matter interaction strength may therefore open a new path to engineer the coherent light-matter coupling in cavity-confined quantum materials.

# SUPPLEMENTARY MATERIAL

The supplementary material contains a full description of the experimental set-up, the thermal evolution of the bare cavity resonance with the procedure employed to correct thermal drifts of the cavity frequency, the full fits of the polariton dispersions across the insulator-to-metal transition performed with the multimode Dicke Hamiltonian, and the estimations of the wave-function components of the phonon-polaritons across the phase transition.

# ACKNOWLEDGMENTS

We acknowledge Professor Dragan Mihailovic (Jožef Stefan Institute, Ljubljana) for providing the 1T-TaS$_2$ single crystals. This work was mainly supported by the European Research Council through the project INCEPT (grant Agreement No. 677488). We acknowledge the support of the Gordon and Betty Moore Foundation through the grant CENTQC.

# AUTHOR DECLARATIONS

## Conflict of Interest

The authors have no conflicts to disclose

## Author contributions

**Giacomo Jarc**: Conceptualization (equal); Data curation (equal); Formal analysis (equal); Investigation (equal); Visualization (equal); Writing-original draft (equal). **Shahla Yasmin Mathengattil**: Data curation (equal); Formal analysis (equal); Investigation (equal). **Angela Montanaro**: Investigation (equal); Writing ‐ review & editing (equal). **Enrico Maria Rigoni**: Writing-review & editing (equal). **Simone Dal Zilio**: Investigation (supporting); Writing-review & editing (supporting). **Daniele Fausti**: Conceptualization (lead); Funding acquisition (lead); Methodology (equal); Writing-review & editing (equal).

# DATA AVAILABILITY STATEMENT

The data that support the findings of this study are available within the article and its supplementary material.

# Supplementary Material for

# Multimode vibrational coupling across the insulator-to-metal transition in 1T-TaS$_2$ in THz cavities

Giacomo Jarc[1,2,3], Shahla Yasmin Mathengattil[1,2], Angela Montanaro[1,2,3], Enrico Maria Rigoni[1,2,3], Simone Dal Zilio[4], and Daniele Fausti[1,2,3,a)].

[1]*Department of Physics, Università degli Studi di Trieste, 34127 Trieste , Italy*
[2]*Elettra Sincrotrone Trieste S.C.p.A., 34127 Basovizza, Trieste, Italy*
[3]*Department of Physics, University of Erlangen-Nürnberg, 91058 Erlangen, Germany*
[4]*CNR-IOM TASC Laboratory, 34139 Trieste, Italy*

*a) Correspondence to: daniele.fausti@fau.de*

# Supplementary Note 1

# Experimental set-up

The full experimental set-up is shown in Fig. S1A. The cryogenic THz cavity is composed of two cryo-cooled piezo-controlled movable mirrors between which 1T-TaS$_2$ is inserted. The movement of each of the two cavity mirrors is ensured by three piezo actuators (N472-11V, Physik Instrumente) with a total travel range of 7 mm and a minimum incremental motion of 50 nm. The independent movement of each of the three piezo ensures the independent horizontal and vertical alignment of the mirrors while the synchronous motion of the three results in a rigid translation of the whole mirror. The tunability of the cavity length sets the frequency of the cavity fundamental mode. Instead, the tunability of the sample position with respect to the mirrors allows to maximize the coupling of the confined cavity field with the targeted excitation. The mirrors are mounted on copper holders, and they are cryo-cooled by means of copper braids directly connected to the cold finger of the cryostat. Since the piezo actuators temperature operational range is 283–313 K, the piezo actuators are thermally decoupled from the mirror supports. The thermal decoupling is realized by placing between the piezo actuators and the mirror holders a PEEK (Poly-Ether Ether Ketone) disk on which the actuators actually act and three ceramic cylinders. These materials are thermal insulators, and they have a low thermal expansion coefficient in the operational temperature range of the cryostat (10 − 300 K). These features ensure the mirrors to be thermally insulated as well as an alignment stability of the cavity in the operational temperature range.

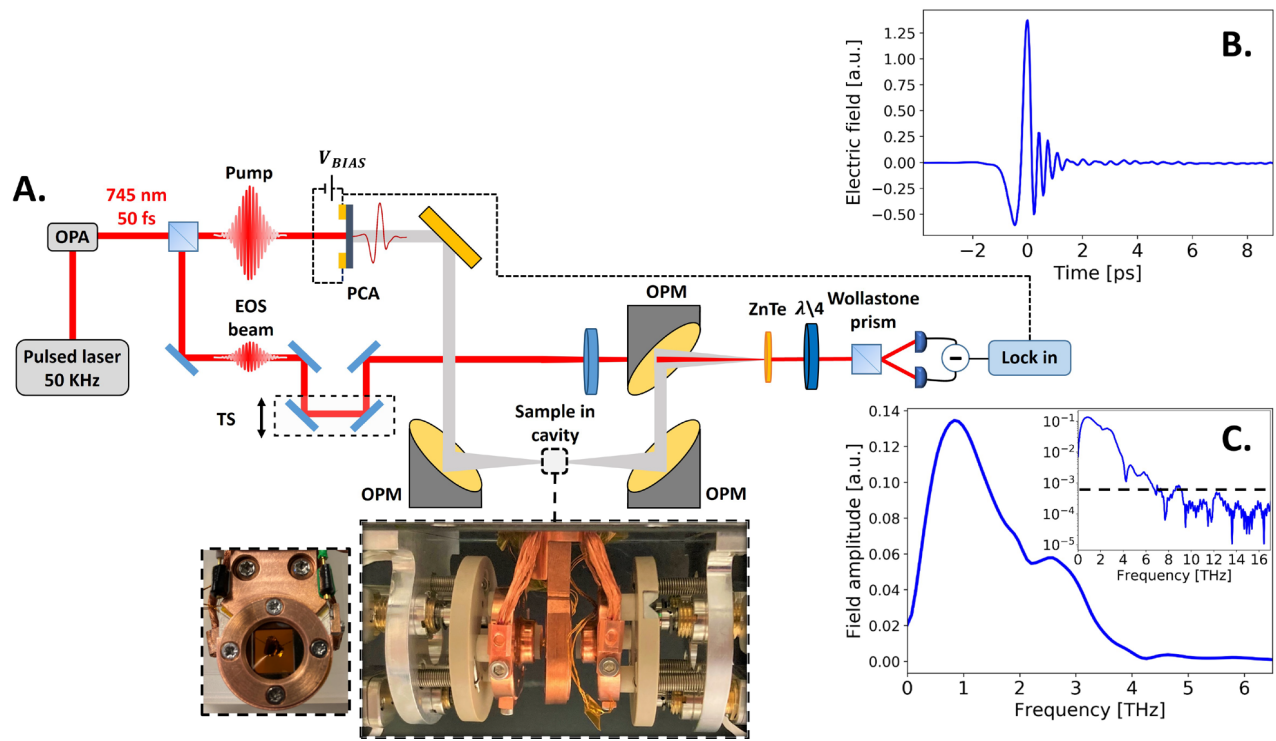

**Figure S1**: **Experimental configuration. A.** Sketch of the ultrafast THz time-domain spectrometer. OPA: Optical Parametric Amplifier, PCA: Photo-Conductive Antenna, TS: Translation Stage, OPM: Off-Axis Parabolic mirror, λ/4: Quarter-wave plate. In the inset, the picture of the tunable cryogenic THz cavity and of the ∼ 10 μm 1T-TaS$_2$ sample mounted within two silicon nitride membranes in the middle of the cavity. **B.** Free-space nearly single-cycle THz pulses generated by the photoconductive antenna (PCA) detected through Electro-Optical Sampling (EOS) in a 0.5 mm ZnTe crystal. **C.** Fourier transform of the nearly-single cycle THz pulses in free space. In the inset, the Fourier spectrum is plotted on a logarithmic axis in order to highlight the spectral content up to approximatively 6 THz. The black horizontal dashed line indicates the noise level.

The chamber, shown in Fig. S1A, is mounted on a flow cryostat, which is supported by a mechanical assembly allowing for the movement of the whole sample in the x, y, and z directions. We stress that the mechanical translation of the sample is particularly crucial for the experiment since it allows us to perform THz transmission measurements of the empty cavity by simply moving the vertical/horizontal position of the whole chamber in Fig. S1A.

The cavity semi-reflecting mirrors are fabricated by evaporating a thin bilayer of titanium-gold (2-15 nm) on a 2 mm crystalline z-cut quartz substrate. The first 2 nm thin layer of titanium are used to increase the adhesion of the following gold layer.

The single crystal ~ 10 μm thick 1T-TaS$_2$ sample is mounted within the two mirrors in a copper sample holder directly connected to the cold finger of the cryostat and sealed between two silicon nitride (Si$_3$N$_4$) membranes (LP-CVD grown) with a window size of 11 x 11 mm$^2$ and a 2 μm thickness. The membranes are supported on a 13 x 13 mm$^2$ silicon frame and do not show any spectral dependence in the THz spectral range employed in the experiment.

We employ broadband THz spectroscopy to characterize the multimode vibrational coupling across the insulator-to-metal phase transition in 1T-TaS$_2$. The layout of the built THz spectrometer based on THz generation in a GaAs-based photo-conductive switcher is shown in the schematic diagram of Fig. S1A.

Ultrashort laser pulses (50 fs pulse duration and 745 nm central wavelength) from a commercial 50 kHz pulsed laser + Optical Parametric Amplifier (OPA) system (Pharos + Orpheus-F, Light Conversion) are split into two to form an intense optical beam for THz generation (6 μJ/pulse) and a weak readout pulse (< 100 nJ/pulse) for time-resolved Electro-Optical Sampling (EOS). Single-cycle THz pulses are generated via the acceleration of the photoinduced carriers in a large-area GaAs-based photoconductive antenna (PCA). The acceleration of the free carriers induced by the pump is achieved by biasing the PCA with a square-wave bias voltage $V_{bias}$ triggered with the laser at a frequency of 1.25 kHz. We employed a biasing square wave with a voltage peak of 8.0 V and a 50 % duty cycle. For an efficient THz generation using 6 μJ pump pulse energy, an area of around 6 mm diameter on the 1 cm$^2$ large emitter is illuminated using a collinear pump beam. The emitted collimated THz beam is then focused on the sample mounted inside the cavity, which is placed in the focal plane of two off-axis parabolic mirrors (OPMs). We estimated the THz spot diameter at the focus position to be ~ 1.5 mm, hence smaller than the lateral dimensions of the 1T-TaS$_2$ crystal (~ 4 mm x 4 mm). The THz field and the readout pulse are then combined and focused on a 0.5 mm ZnTe crystal, which acts as the electro-optical crystal. After the electro-optical crystal, the probe beam, variable delayed in time through a translation stage (TS), is analyzed for its differential polarization changes induced by THz in the ZnTe crystal, which maps the time evolution of the ultrafast THz field. This is carried out by standard Electro-Optical Sampling (EOS), by splitting the two probe polarizations with a Wollaston prism and measuring the differential intensity recorded on a pair of photodiodes. The resulting differential signal is then detected using a lock-in amplifier (SR830, Stanford Research System) referenced at the frequency of the bias voltage ($V_{bias}$). The entire system is purged with nitrogen to eliminate THz absorption coming from the water vapour in the ambient atmosphere. We show in Fig. S1B the measured electric field of the generated THz pulse and its calculated Fourier spectrum (Fig. S1C). As shown, the input field is, indeed, a nearly single-cycle THz pulse with the spectral content reaching ~ 6 THz, as highlighted in the logarithmic scale plot in the inset of Fig. S1C. We estimate the signal-to-noise of the detected THz field to be 4.6 x 10$^4$ with a temporal phase stability ≤ 30 fs.

# Supplementary Note 2

# Thermal evolution of the bare cavity resonance

As presented in the main manuscript, the employed THz cavity allows for performing temperature-dependent studies of the cavity-confined material. In this sense, it is crucial to characterize how the bare cavity properties evolve with temperature to subsequently correct possible thermal drifts.

Fig. S2A presents the thermal evolution of the empty cavity transmission measured in the range 80-280 K. A thermal drift of the cavity resonance is detected. The thermal drift is quantified in Fig. S2B where we present the thermal variation of the cavity length ΔL estimated from the cavity transmission resonances. Overall, a length variation of ∼ 65 µm is detected within the employed temperature range, consistent with a linear thermal expansion of the copper mirrors mounts. Importantly, as shown in Fig. S2C, the thermal drift of the cavity length does not significantly affect the quality factor of the cavity. This implies that the cavity alignment, which in turn sets the Q-factor of the Fabry-Pérot resonator, can be considered temperature-independent.

Given these evidences, we schematic present in the following the procedure to account for thermal drifts of the cavity resonance which, obviously, become more important at higher cavity frequencies.

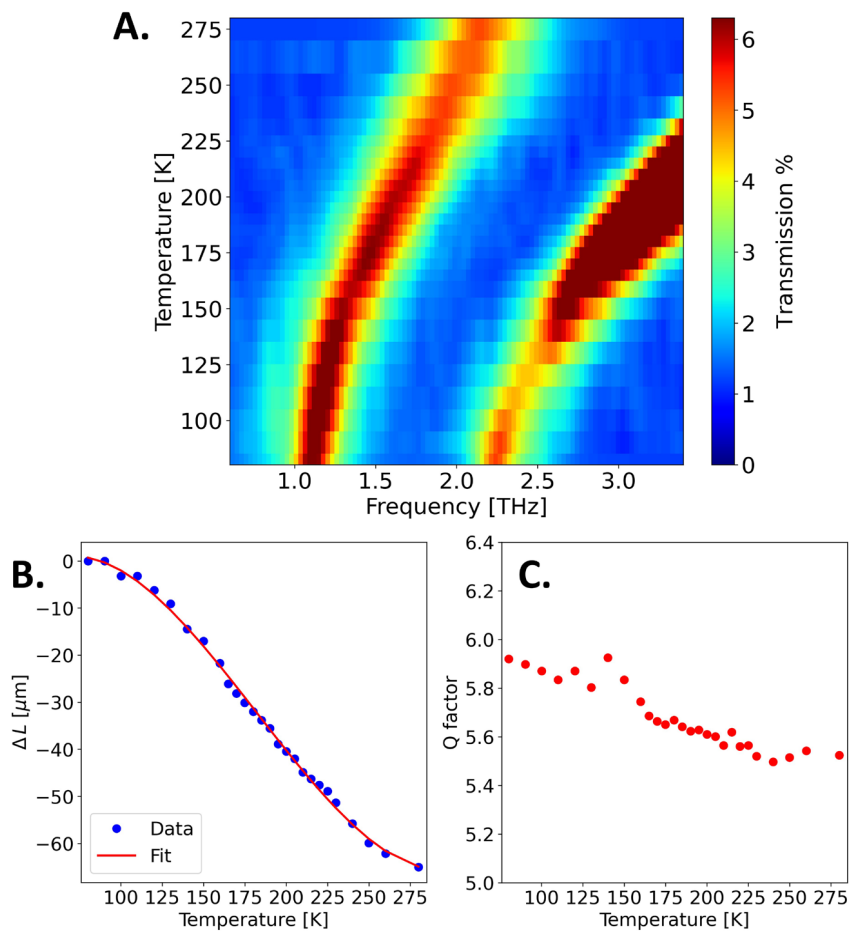

**Figure S2**: **Temperature evolution of the empty cavity properties. A.** THz transmission of the empty cavity as a function of the temperature (measured on the sample's mount). **B.** Thermal shift of the cavity length (ΔL) estimated from the THz transmission presented in A. The corresponding polynomial fit is shown in red solid line. **C.** Quality factor of the empty cavity as a function of the temperature calculated from the THz transmission in A.

The employed procedure for correcting this thermal drift and subsequently obtain temperature scan at a fixed cavity mode is the following:

i. We set the frequency $\omega_c$ of the bare cavity at 80 K, by tuning the distance of the cavity mirrors so that a transmission peak is observed at the desired frequency $\omega_c$.

ii. For the cavity length set at 80 K, we move the THz beam on the sample and measure the temperature-dependent THz transmission of the coupled cavity.

iii. For each spectrum at fixed temperature obtained in ii), we assign an effective cavity length $L_{eff}$. The latter can be estimated by fitting the thermal variation of the cavity length $\Delta L$ for a representative cavity resonance (see the polynomial fit of Fig. S2B). With this procedure, each cavity spectrum can be labelled with two indices: its temperature and its effective cavity length $L_{eff}$.

iv. We collect all the cavity spectra sharing the same effective cavity length $L_{eff}$ and construct the corrected temperature scan, in which the thermal drift of the cavity resonance shown in Fig. S2A is removed.

# Supplementary Note 3

# Extended fits of the polariton dispersion across the insulator-to-metal transition

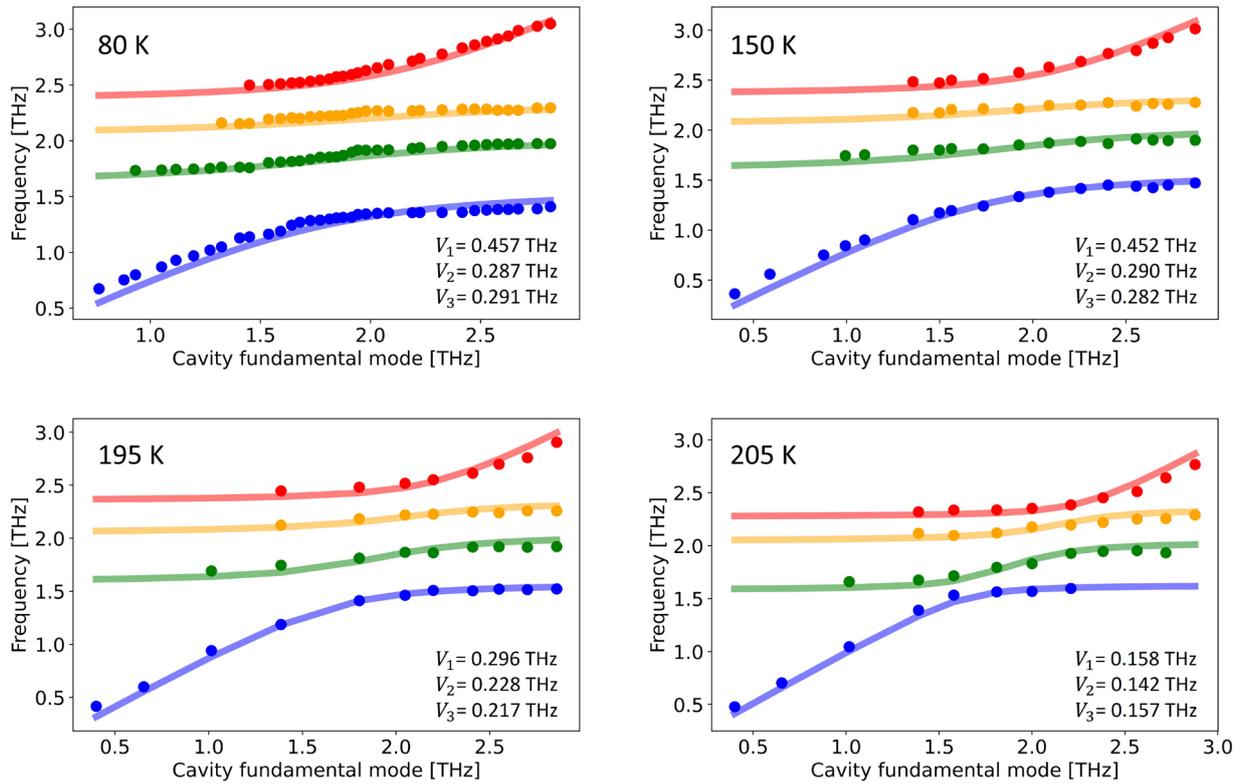

**Figure S3**: **Fits of the vibro-polariton dispersions across the insulator-to-metal transition.** Fit of the THz vibro-polariton branches as a function of the temperature (indicated inside the plots). The fits of the polariton dispersions have been performed approaching Tc (~ 215 K) from the insulating phase, where the multimode polariton modes are visible. For each temperature, the circles correspond to the measured polariton peaks. The solid curves represent the eigen-values obtained from the multimode coupled oscillator model (Eq. 1, main manuscript) fitted to the measured transmission peaks. Note that only peaks above the noise level have been fitted. The fitted light-phonon collective couplings within the cavity $V_{i(i=1,2,3)}$ are indicated in legend for each temperature.

# Supplementary Note 4

# Wave-function components of the polariton states across the insulator-to-metal transition

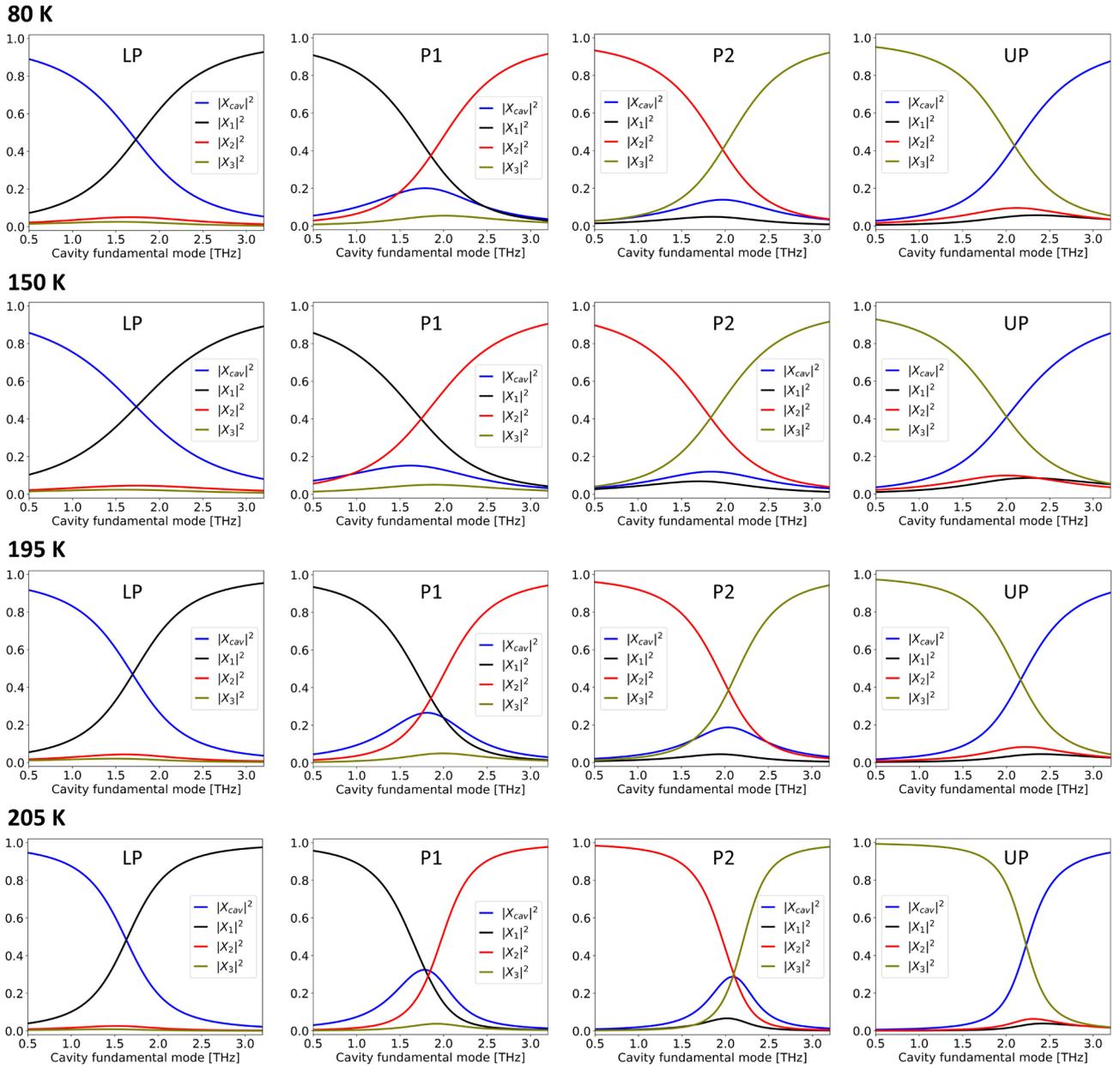

**Figure S4**: **Wave-function components of the phonon-polariton states across the insulator-to-metal transition.** Cavity ($|X_{cav}|^2$) and phonon ($|X_{1,2,3}|^2$) fractions of the four polariton (LP, P1, P2, UP) wave-functions as a function of the cavity mode across the insulator-to-metal transition. Estimations are based on the coupled oscillator model (Eq. 1, main manuscript) employed to fit the dispersion of the polariton peaks at different temperatures. Except of a tiny increase of the cavity fraction of the middle polaritons (P1, P2) around 205 K, the trends of the phonon-polariton wave-functions as a function of the cavity frequency do not exhibit a significant modification approaching the metallic phase.